# Clearing the Fog of War: Lessons from Vibe Coding a Red Alert Map Hacker

Yunfei Hou[1], Xiangyu Li[1], Yue Dong[2], Jia Chen[2], Evangelos Papalexakis[2]
[1] California State University San Bernardino, San Bernardino, United States
hou@csusb.edu, xiangyu.li@csusb.edu
[2] University of California Riverside, Riverside, United States
yue.dong@ucr.edu, jia.chen@ucr.edu, evangelos.papalexakis@ucr.edu

***Abstract*—AI coding agents are changing what students need to know about programming: less time may be spent on syntax and grammar, but more practice is required to frame problems, review generated code, and make ethical design choices. This paper presents a classroom-ready example for teaching vibe coding: creating a map hacker for the Command & Conquer: Red Alert game that reveals every unit on the map. Built with an AI coding agent and the game's open-source code, we use the project to highlight the strengths and limits of vibe coding. The AI agent was effective for scoped implementation, debugging, refactoring, explaining unfamiliar APIs, and interpreting low-level machine code and memory data. It was much less reliable on open-ended goals that required domain knowledge, problem decomposition, and human judgment. We also report formative classroom evidence from two sections of a senior capstone course: students introduced to vibe coding performed similarly on easier lab assignments but completed the most demanding lab at substantially higher rates. At the same time, student uptake was uneven, raising equity concerns because motivated students benefited most. We conclude with adoption guidance for computing instructors: introduce AI-assisted development early, keep computational thinking central, teach version control, and make responsible AI use and ethical hacking explicit learning goals.**

## I. Introduction

For most of computing's history, writing software has been an artisan craft. Much as a weaver once worked a hand loom, a programmer typed each section of code by hand, one line at a time. We could call this *Artisan Coding*: the traditional process in which people manually write, debug, and document software without AI-generated assistance. This notion is not merely nostalgic: programmers have long taken pride in the artistry of their work, from Knuth's characterization of programming as an art [1] to computing educators who place the beauty and joy of coding at the center of the classroom [2]. With the rise of AI coding agents, programming is undergoing its own industrial revolution, mechanizing much of that hand work and changing what it means to write software. As educators, our aim in this paper is to explore how to help students keep pace with this shift.

In this paper, we present a classroom-ready example for learning vibe coding: building a map hacker for Command & Conquer: Red Alert that reveals the positions of all units in this classic real-time strategy game. We also report on students' perceptions on their vibe coding experience with observations from a senior capstone course. We conclude with reflections on teaching and learning programming in the age of AI-augmented development.

Our central reflection is that AI shifts the work of programming. While building the map hacker, the AI agent excelled at well-scoped tasks such as implementation, debugging, and explaining large legacy codebase. It struggled, however, with open-ended goals where domain knowledge and human judgment were decisive. We therefore argue that programming courses, from the introductory level onward, should introduce vibe coding thoughtfully while keeping computational thinking central, adopting version control early, and treating the ethics of AI-assisted work as a first-class concern.

We also see vibe coding as an opportunity to broaden participation in computing, and the shock it brings to the job market is temporary. History suggests that automating a craft often grows the field around it: the power loom displaced many weavers, yet the textile industry expanded dramatically; the camera ended most portrait painters' livelihoods, yet painting flourished as an art form. We expect AI to transform the work of programmers in similar ways, lowering the cost of building software and bringing computing to far more people and problems than before. Education should therefore prepare students not to compete with AI at writing code, but to leverage those productivity tools with computational thinking, domain knowledge, and ethical judgment to solve practical problems.

## II. Background and Related Work

### A. Vibe Coding and AI-Assisted Programming

AI-assisted programming has shifted from line-by-line autocompletion toward generating whole programs from natural-language intent, a practice now popularly called *vibe coding*. The term was coined by Andrej Karpathy in early 2025 to describe a workflow in which the developer states what they want in a prompt, the large language model (LLM) writes the code, and the developer steers through further prompts while paying little attention to the code itself [3]. This lowers the barrier to building software, but it raises a question for education: if an LLM can produce working code from a description, what should students learn, and how do they retain the conceptual understanding that hand-writing code once helped them to acquire?

### B. Student Learning with AI Assistance

Effective AI use still depends on students' ability to formulate, interpret, and validate code. Early studies showed that AI could solve introductory programming tasks [4], prompting proposals to emphasize code evaluation, debugging, and revision in learning activities [5], and to update learning objectives and assessments to support critical AI use while retaining programming fundamentals [6]. Bhatt et al. [7] examined students' experiences with ChatGPT as a programming tutor. Students differ in recognizing unhelpful suggestions and monitoring their understanding [8]. Gardella et al. [9] found higher immediate task performance with Copilot than with a human programming partner, but no significant difference between conditions on an individual retest one week later. Higher scores on a foundational CS knowledge assessment were associated with better performance on GUI-oriented vibe-coding tasks [10], while inaccurate or unclear AI explanations complicate code comprehension [11]. These findings make students' reasoning and independent understanding central to evaluating what they learn through AI-assisted programming.

### *C. Teaching Programming with AI Assistance*

Computational thinking remains essential throughout design, integration, and revision. Reihanian et al. [12] identify graduated hints and feedback grounded in student work as ways to support learners' reasoning. Pădurean et al.'s Prompt Programming platform provides practice in iterative prompting, coordinating multiple functions, and evaluating generated code through execution [13]. At the course level, Bardach and Murrah connect debugging and testing exercises to a staged full-stack project, requiring students to defend design choices and write architecture reports without AI assistance [14]. Studies of larger projects extend this emphasis. Wang et al. highlight requirements, design, systematic testing, and code review in an upper-level course and a separate developer-led implementation [15]. Oliveira et al. examine AI feedback and subsequent revisions within capstone GitHub workflows [16]. These approaches make problem formulation, decomposition, and evaluation visible throughout development. They suggest organizing AI-supported assignments around decisions students must explain and verify, with development artifacts providing evidence of their reasoning alongside the completed software.

Building on these instructional approaches, we present an AI-assisted map-hacker project using Electronic Arts' popular, open-source Red Alert codebase [17]. The project could serve as engaging courseware for exploring the strengths and limitations of vibe coding. Its read-only, single-player design also supports discussion of the ethical boundaries of hacking. Alongside this development case, we draw observations from a senior capstone course to examine how AI assistance changes the knowledge and judgment students need to guide software development. Together, the project and classroom observations motivate teaching practices that keep human judgment central , while evaluating the map-hacker activity's effects on student learning remains future work.

## III. The Red Alert Map Hacker

### *A. About the Game: Command & Conquer: Red Alert*

The original Command & Conquer: Red Alert was developed by Westwood Studios in 1995. The game helped define the real-time strategy genre: each player gathers resources, builds units, and fights while the enemy does the same. In 2020, Electronic Arts took the original game, updated its graphics, and rebuilt its multiplayer features, releasing the Command & Conquer Remastered Collection on the Steam platform. Later that year, key game components, including the game logic and map editor, were released on GitHub, with support for user modding and Steam Workshop integration. Modding lets players customize or expand a game, and Steam Workshop is the Steam platform's built-in hub for finding, sharing, and installing those customizations.

### *B. Why Red Alert Is a Good Vibe-Coding Project*

Red Alert is a good fit for experimenting with vibe coding for several reasons. First, it is a practical project: the GitHub repository is well received, with over 21k stars, and contains about 1,200 files; too many to study manually, but a good size for AI-assisted code exploration. Second, it combines old and new game design: the original game logic was preserved, while the remastered version uses a modern game engine, GlyphX, for its user interface. The game integrates with the Steam platform's modern networking (to support multiplayer) and Workshop (to support user customization) features. Third, the game is fun and easily engages students, while hacking it provides a practical example that leads to discussions of ethics in programming.

### *C. Building the Map Hacker*

In this section, we present the design of the map hacker and describe potential learning outcomes of using it as an example for vibe coding. The project repository is Red Alert Map Hacker [18]. The open-source map hacker works only for single-player games, not multiplayer games, to avoid interfering with fair play. It is written in Python 3.13 (Anaconda release) using Claude Code with Opus 4.7.

In Red Alert, the area a player has not yet explored is covered by shroud, also known as the fog of war. The goal of a map hacker is to remove this shroud and reveal all units on the map. Refer to Fig. 1.

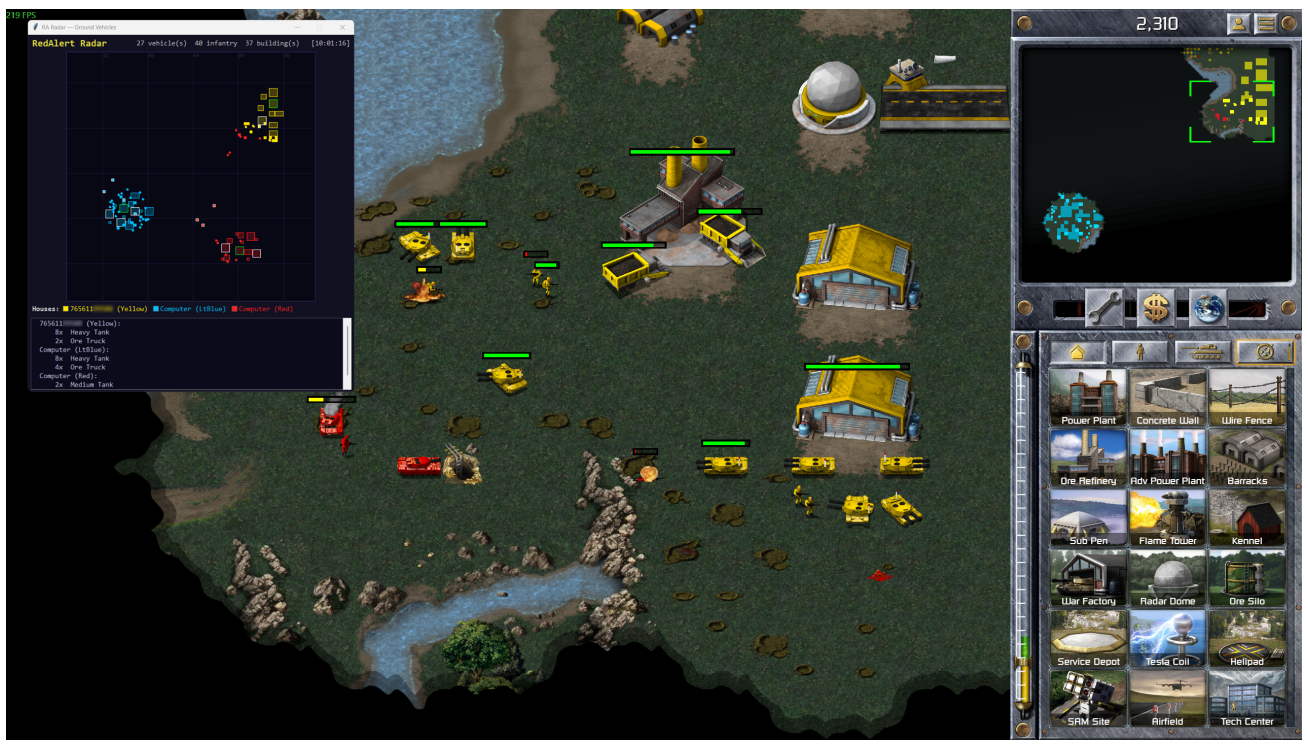

Fig. 1. Screenshot of the Map Hacker (on top left) during a game.

*1) Showcasing Vibe Coding: A GUI from a Prompt:* As a demonstration of the power of vibe coding, we can begin by building a graphical user interface from just a few lines of prompts. Our GitHub page hosts two versions of the map hacker. The first is a command-line app that preserves traces of the incremental development of the map-revealing logic; the second draws a mini-map in a separate window, which looks similar to the radar map in the game. The mini-map version can be created quickly from the command-line source code, and students can explore how to change its display settings. For example, highlighting units or showing player information on Money, Power, Unit Count etc., by prompting in casual language, without writing code.

*2) Understanding the Game Logic:* Our map hacker works by reading the game process's memory to find the byte signatures and structural patterns that store units' positions. It only reads memory, never writes any changes, and then reports or draws the unit positions on its own interface. This is the point where human judgment plays a key role: programmers must make a design choice based on an understanding of game logic. At least two other directions might work: 1) Study how shroud-related information is stored and rendered, then change the corresponding values in the game's memory. This approach is less stable, because shroud state is dynamically allocated on the game process's

This work was supported in part by the National Science Foundation under Awards 2524227 and 2524228.

heap and changes between runs; writing to the wrong address will crash the game. 2) Capture the messages about players' actions, such as building or moving units etc., that pass between the UI and the network. This approach is harder, because the UI and network are built on lower-level operating-system support: there is a semantic gap, and capturing the raw data at that level does not help us interpret it back into the data structures the game uses. During our tests, when we let the AI try these two methods, or directed it to freely explore other options, it would keep trying without a clear goal (seemingly losing focus) or arrive at wrong conclusions. This can serve as an example that cautions students against over-relying on AI to solve problems.

*3) Decomposing the Problem into Trackable Steps:* The AI tool was effective at giving an overview of the existing game design from the developers' source code. With a few prompts, such as how information about players' units is stored or how the coordinate system works, the AI tools can help quickly decide on the map hacker's logic. Our key observations from the game's source code are as follows:

The game logic (located in redalert.dll) stores all game objects in global, pre-allocated, fixed-size object pools ( i.e., heaps of contiguous arrays with self-describing headers), one pool per category for houses (e.g., a player's name, color and country faction), ground vehicles (e.g., a tank's position and hit points), infantry, buildings, vessels, and aircraft. This layout is the key enabling feature of the map hacker: each pool can be located by a structural-signature scan of its header, and once its buffer pointer is found, every slot is reachable at a fixed stride (in bytes).

The global variable structure is organized by role, as shown in Listing 1. The object hierarchy, organized by class derivation, is shown in Listing 2. The house pool (`Houses = TFixedIHeapClass<HouseClass>`) holds one record per player, with the player's name, available money, and per-category unit counts. Each live-object pool (e.g., `Units = FixedIHeapClass<UnitClass>`) holds one record per object, carrying the object's type index (a handle into a separate table of type definitions that supplies the human-readable name), its map coordinate, owning-player index, and hit points. Joining a live object's type index against the type table, and its owner index against the house pool, yields a fully-labelled such as "Mammoth Tank at (45,72) owned by USSR" that the map hacker reports.

The next step is to find those global object pools, but they are dynamically allocated on the heap, so their addresses are not fixed. The traditional way to locate such a value is by **perturbation**: observe a value in the game such as money, scan memory for that value, change it in-game (e.g., build a unit), then rescan for the new value, narrowing the candidates until the single address holding the money variable is found. This process may have to be repeated across several runs of the game, because the data's absolute address moves between runs or builds, i.e., an address that works in one session may be invalid in the next.

Because the absolute address is unstable, a common approach in game hacking is to anchor on the code that accesses the value rather than on the value itself. The machine-code instruction that reads or writes money lives at a relatively stable location inside the module, and the bytes surrounding it form a recognizable fingerprint. That fingerprint is also known as an Array of Bytes (AOB) signature: a scanner searches the module's memory for the exact byte sequence to relocate the instruction even after the module is reloaded at a different base address.

```
TFixedIHeapClass<T>        (HEAP.H — the pool template; one pool per type in GLOBALS.CPP)
│
├── live-object pools     (one slot per object on the map)
│   ├── Units      = TFixedIHeapClass<UnitClass>      ← ground vehicles (stride 587)
│   ├── Vessels    = TFixedIHeapClass<VesselClass>    ← naval
│   ├── Infantry   = TFixedIHeapClass<InfantryClass>  ← infantry (stride 550)
│   ├── Aircraft   = TFixedIHeapClass<AircraftClass>  ← aircraft
│   └── Buildings  = TFixedIHeapClass<BuildingClass>  ← buildings (stride 436)
│
├── owner table
│   └── Houses     = TFixedIHeapClass<HouseClass>     ← player/AI records (one slot per player)
│
└── type tables         (read-only unit and building definitions)
    ├── UnitTypes      = TFixedIHeapClass<UnitTypeClass>
    ├── VesselTypes    = TFixedIHeapClass<VesselTypeClass>
    ├── InfantryTypes  = TFixedIHeapClass<InfantryTypeClass>
    ├── AircraftTypes  = TFixedIHeapClass<AircraftTypeClass>
    └── BuildingTypes  = TFixedIHeapClass<BuildingTypeClass>
```

**Listing 1:** Global object pools (TFixedIHeapClass), organized by role.

```
AbstractClass                     (identity + position — RTTI, Coord)
└── ObjectClass → MissionClass → RadioClass   (map presence, HP, orders, radio)
    └── TechnoClass               (owner, weapons, armor — has House)
        │                         (+ mix-ins: Flasher/Stage/Cargo/Door)
        ├── FootClass                 (can move across the map)
        │   ├── DriveClass            (wheeled/tracked ground movement)
        │   │   ├── UnitClass         ← ground vehicles
        │   │   └── VesselClass       ← naval
        │   ├── InfantryClass         ← infantry (sub-cell foot movement)
        │   └── AircraftClass         ← aircraft (includes FlyClass)
        └── BuildingClass             ← buildings (stationary; no FootClass)
```

**Listing 2:** Object class hierarchy by class derivation.

A typical AOB is shown in Listing 3 (credited to the user Recifense, from fearlessrevolution.com). It anchors on the `HouseClass` accessor that computes the player's available money. Disassembling these bytes gives the operations shown in Listing 4.

```
8B 81 B4 01 00 00 03 81 B0 01 00 00 C3 CC CC
?? ?? ?? ?? ?? ?? ??
```

**Listing 3:** Byte signature for the money accessor.

It loads `Credits` from offset `0x1B4`, adds `Tiberium` from offset `0x1B0`, and returns the sum in `eax`. When the scanner finds this byte sequence, it has located the exact code that computes the player's money, and because `ecx` already points at the player's `HouseClass` when the function runs, the offsets `0x1B4` and `0x1B0` yield `Credits` and `Tiberium` directly, with no need to know the absolute address in advance.

```
8B 81 B4 01 00 00  mov eax,[ecx+000001B4]
; eax = Credits
03 81 B0 01 00 00  add eax,[ecx+000001B0]
; eax = Credits + Tiberium
; Result: Available_Money
C3  ret
; Return eax to the caller
CC  CC
; int3 padding between functions
?? ?? ?? ...   ; Wildcard bytes
```

**Listing 4:** Disassembly of the money-accessor bytes.

It is interesting to note how money is represented in the game. `Available_Money = Credits + Tiberium` that is cash on hand plus the value of Tiberium ore (a fictitious resource) stored in refineries or silos. Students who scan for a single 'money' value may find either the Credits field or the Tiberium field, but not both. This makes a good example for showing students why domain knowledge and clear naming matter when writing prompts. In our early prompts, asking the AI tool to search for 'money' did not work well because terms such as money, credits, cash, gold, and gems were treated as interchangeable. The AI tool became more useful only after money was defined precisely as the sum of Credits and Tiberium.

*4) A Surprising Find: AI-Derived Structural Signatures:* When the AI was asked to find an AOB in memory to locate the global unit-list pools (including the live Houses, Units, Infantry, and Buildings heaps), it adopted a different strategy that worked surprisingly well. It studied the definition of `HouseClass`, computed its memory layout, and then located the pool by validating the expected initial values. In other words, it used a signature derived not from observed bytes but from the *source-level semantics* of the structure.

The AI agent began with a structural fact rather than an opcode signature: every game-object pool is an instance of the template `TFixedIHeapClass<T>`, so all pools share one common header. It studied `HouseClass` and computed the 24-byte header layout. The derived layout of the `TFixedIHeapClass<HouseClass>` header (24 bytes) is shown in Listing 5.

From this, the AI derived a "byte signature": four consecutive 32-bit integers satisfying these value constraints, followed by a non-null pointer. It also recognized that the number of house slots is fixed at 21, because the `TotalCount` of the `Houses` pool is set to `HOUSE_MAX = 21` at construction. An example of an actual `Houses` header is shown in Listing 6.

Here `15 00 00 00` is decimal 21, the `HOUSE_MAX` anchor used to locate potential candidates. The AI agent also performed additional validations, which seem somewhat redundant: range checks on `Size` (4000 < Size < 50000), `ActiveCount` (1..21), then validated each candidate by dereferencing `Buffer` and confirming that a slot held a printable-ASCII house name.

The global unit-list pools for infantry, ground vehicles, and buildings are found in a similar way: the header shape locates candidate heaps, and the per-slot RTTI byte is the discriminating signature that confirms which heap is which. The RTTI byte is Red Alet's custom version of the Run-Time Type Information (RTTI) mechanism; it sits at the root of every game object and uses a 1-byte type ID that indicates which C++ class an object belongs to (e.g., `UnitClass`, `InfantryClass`, or `BuildingClass`).

## IV. Classroom Context

To examine vibe coding in a real course, we compared two sections of a senior capstone engineering design course, taught by the same instructor in consecutive semesters. Both sections completed the same three lab assignments of increasing difficulty: Lab 2, a RESTful web server for JSON data exchange; Lab 3, an Android app with multiple Activities; and Lab 4, a combined web server and Android app that exchange JSON data between them. Lab 4 was the most demanding of the three. One section (n = 35) was introduced to vibe coding using ChatGPT-4o; the other (n = 30) completed the labs without AI assistance. Because the two cohorts were taught in different semesters, this comparison is suggestive rather than

```
offset  field          size  expected value
+00     vtable         4     Unconstrained pointer
+04     IsAllocated    4     Bitfield container (unconstrained)
+08     Size           4     sizeof(HouseClass)   — range: 4000-50000
+0C     TotalCount     4     21                   — HOUSE MAX
+10     ActiveCount    4     1 to 21              — current players
+14     Buffer         4     pointer to slot[0]   — must be ≥ 0x10000
```

**Listing 5:** Derived 24-byte layout of the TFixedIHeapClass<HouseClass> header.

```
B0 4F 8A 5F  00 00 00 00  D8 1D 00 00  15 00 00 00  03 00 00 00  E0 50 8C 5F
└─ vtable ─┘ └─ alloc ──┘ └ Size=7640 ┘ └Total=21 ┘ └Active=3 ┘ └─ Buffer ─┘
```

**Listing 6:** Example Houses header matching the derived byte signature.

controlled: differences in student composition and semester effects may also contribute to the differences we report.

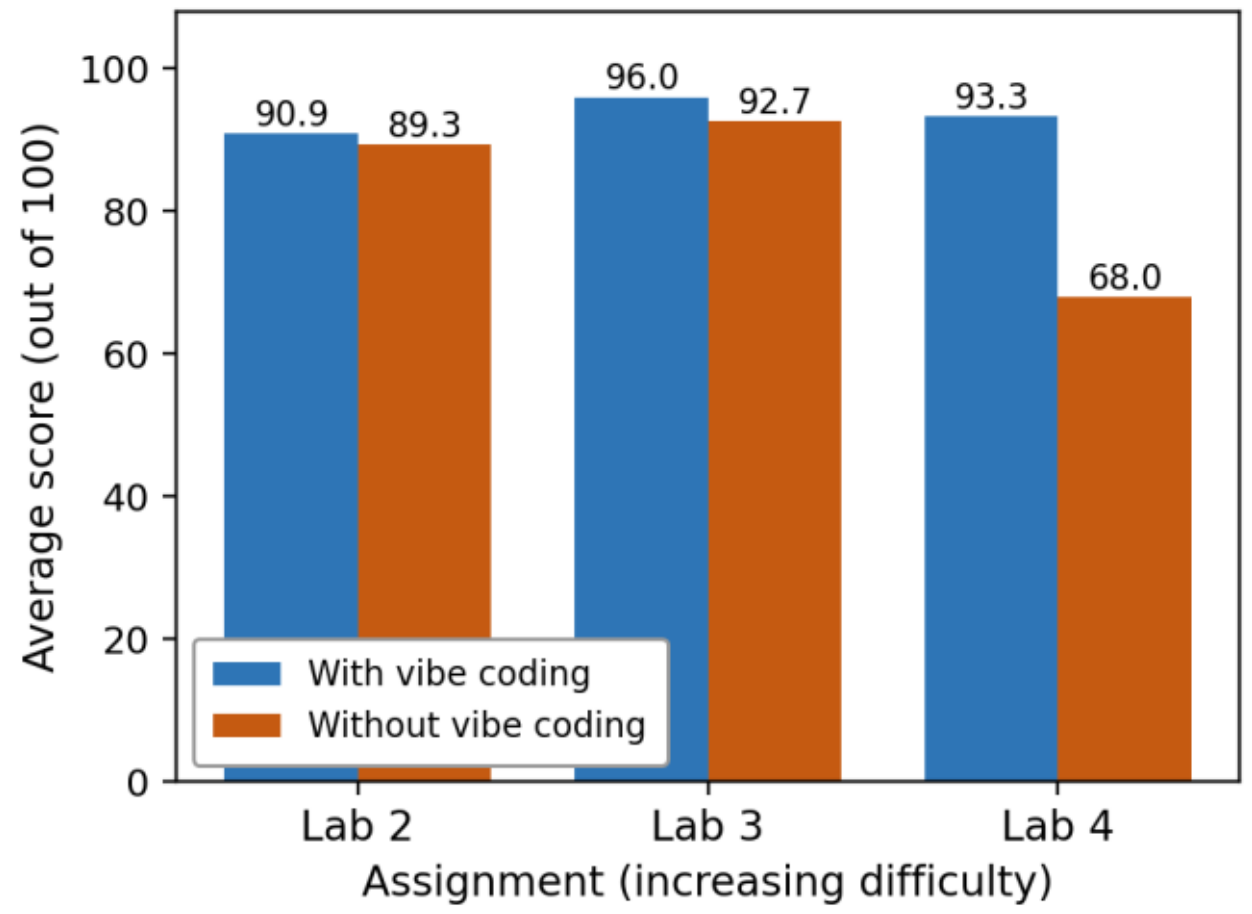

Fig. 2. Average lab scores for the two course sections (n = 35 with vibe coding, n = 30 without).

Fig. 2 shows the average score on each lab for the two sections. The groups are comparable on Labs 2 and 3 but diverge sharply on the most demanding assignment, Lab 4. On the two easier labs, the groups performed comparably (90.9 versus 89.3 on Lab 2, and 96.0 versus 92.7 on Lab 3). On Lab 4, the gap was large: the vibe-coding section averaged 93.3, while the section without it averaged 68.0. The difference came mostly from students who did not complete Lab 4 at all, as zero scores were far more common without vibe coding. In short, on the most demanding assignment, which required students to build and integrate a web server and a mobile client, AI agent helped more students finish. AI agents remove much of frictions on syntax, APIs and compiler errors, the challenges then move toward higher-level work: formulating problems precisely, decomposing systems into components, judging whether generated solutions are correct, and integrating subsystems into a coherent whole.

The reception of the tools, however, was uneven. Most students welcomed vibe coding, but not all shared that enthusiasm. We observed a widening gap: already-motivated, high-performing students used the tools to do more, while some struggling students engaged with them little and risked falling further behind. Prather et al. similarly report uneven benefits and metacognitive difficulties among novice programmers [8], although their study and our capstone observations concern different populations and tasks. These observations raise concerns about unequal benefits from AI assistance. Instructors should explicitly teach students how to evaluate, test, and reason about AI-generated code, rather than assuming that access to the tools alone will benefit all learners equally.

## V. Reflections on Learning Outcomes

Building the map hacker was as much a motivating game project as a programming exercise with AI, and that combination is what makes it useful as a teaching tool. In this section, we discuss what the project taught us: where AI helped, where it misled us, and what students should learn for themselves.

### A. What AI Did Well

While building the map hacker, vibe coding worked best when a task was well scoped and could be described precisely. The AI agent was strongest at:

1) Programming from clear, specific descriptions. When we could state exactly what we wanted, such as scanning the heap for the Infantry list much as we had already done for Ground Vehicles, the AI produced working code in seconds. This reinforced the point that understanding the business logic is crucial.

2) Debugging, by reading error output and proposing targeted fixes. Pasting a crash or a stack trace usually yielded a correct diagnosis faster than manual inspection. The AI tool also worked very well at interpreting assembly opcodes and raw memory contents.

3) Explaining architecture and APIs, and locating the right syntax in large documentation. The agent summarized the relevant parts of the 1,200-file game codebase and pointed us to the classes and calls that mattered. This let students orient quickly in a large, real codebase instead of stalling on where to begin.

4) Refactoring and code review. As the map hacker grew from a command-line probe into a mini-map app, the agent restructured it quickly on request. Industry often resists refactoring because it brings no immediate product value, but lowering its cost makes it a more realistic habit to teach.

### B. Pitfalls and Over-Reliance

The same fluency introduced risks, especially for beginners. We observed two recurring drawbacks:

1) Waiting for AI responses can interrupt the programmer's train of thought. During long memory-scanning experiments, the pauses between prompts repeatedly broke our concentration on the problem. Taking notes and running several agents in parallel helped, but a smoother workflow is still needed.

2) Over-reliance on AI can erode understanding of the big picture. When we let the agent freely explore how to defeat the shroud, it churned through plausible but wrong approaches without converging. Vibe coding is especially hard for novices, who may follow such detours and lose control of the design.

### C. Ethics and Responsible Hacking

Hacking a game is a natural hook for discussing the ethics of software. Our map hacker only reads memory and runs only in single-player mode, so it never alters the game or disrupts fair online multiplayer play; making this constraint explicit lets students weigh curiosity-driven reverse engineering against cheating, and reading a system against tampering with it. Because the underlying source is released under an open-source license [17], the example also opens a discussion of what such a license does and does not permit.

### D. Implications for Introductory Programming

More broadly, although our evidence comes from a senior capstone, the benefit was greatest where the programming was hardest, which argues for building these programming skills early. We therefore propose introducing vibe coding from introductory courses onward, guided by the following principles:

1) Keep computational thinking central. Domain knowledge and business logic, not syntax, drive the design, and they are what students must supply to steer the AI. The map hacker worked only after we understood how the game stored its objects, a judgment the tool could not make for us.

2) Introduce version control (e.g., Git) early, so students can track changes and explore alternatives quickly. We relied on frequent commits to compare different problem-solving strategies and to recover when an experiment broke the build. Version control becomes more valuable when an AI is generating frequent changes.

3) There is no need to rush to adopt every new tool or agent; let them mature over time. We saw tools change rapidly even over this single project. For example, effectiveness often dips after Claude Code auto-compacts its context, though better use of project memory file (i.e., Memory.md feature) helps alleviate the problem.

4) Helping others set up and configure AI agents, drawing on domain knowledge and business sense, may itself become a valuable skill. This requires computing students to broaden their knowledge and become domain experts.

## VI. Conclusion and Future Work

We presented a vibe-coding example, a Red Alert map hacker that reveals every unit by reading game memory, and used it to examine what AI-assisted programming changes for teaching programming. The project shows AI at its best on scoped implementation, debugging, refactoring and code comprehension, and at its weakest on open-ended goals that demand human judgment, domain knowledge, and careful problem decomposition. These observations are drawn from the authors' development experience and from a suggestive course comparison. The central direction for future work is to deploy the example more widely, including in introductory courses, and measure its effect on student understanding, engagement, and ethical reasoning.

Beyond the map hacker, the open-source release invites projects that improve the game rather than exploit it. We encountered several natural next steps. First, the lock-step multiplayer model lags once unit counts approach the cap (for example, 500 tanks): because every peer must simulate every unit each frame and stay perfectly in sync, both the per-frame work and the state exchanged at each synchronization point grow with the number of objects, and the periodic keyframes appear to carry more per-unit data than the connection can pass smoothly. Second, the rule-based computer player could be made considerably smarter, with better build orders, scouting, and strategy that adapts to the opponent. Third, the replay system supports only basic playback and could adopt features expected from modern real-time strategy games, such as variable-speed seeking, per-player build-order and economy timelines, army-value and actions-per-minute statistics, and the ability to follow any player's view or watch with the fog of war removed. Finally, long-standing bugs from the original design remain, including an attack-range calculation that gives the Tesla coil extra reach when firing from south to north, and pathfinding that lets units bunch up and stick together. Each of these is a well-scoped and ethically clear project in which AI assistance could accelerate real improvements to a much-loved game.